\documentclass[10pt,preprint,two column]{aastex}
\begin{document}

\title{Diffuse Ionized Gas in irregular galaxies: I- Gr 8 and ESO 245-G05 }
\author{A.M. Hidalgo-G\'amez }
\affil{Instituto de Astronom\'\i a, UNAM, Ciudad Universitaria, Aptdo. 70 264,
C.P. 04510, Mexico City, Mexico\\ and\\ Escuela Superior de F\'{\i}sica y Matem\'aticas, IPN, U.P. Adolfo L\'opez Mateos, Mexico City, Mexico}

{\bf Abstract}

We have studied the spectral characteristics of the Diffuse Ionized Gas in 
two irregular galaxies with low metallicities and intermediate Star 
Formation Rates: ESO 245-G05 and Gr 8. The [OIII]/H$\beta$ in these galaxies 
is higher than in the DIG of spiral galaxies but not as high as in 
other irregular galaxies previously studied,  such as IC 10 and NGC 6822. 
The [NII]/H$\alpha$ and [SII]/H$\alpha$ ratios have very small values, indicating the absence of shocks as the ionization source for this 
gas. This ionization can be explained in both galaxies with photon leakage 
from the H\,{\sc ii} regions as the only source. The percentage of 
photons escaped from the H\,{\sc ii} regions is small in ESO 245-G05, of only  $35 \%$, but varies from $35 \%$ up to $60 \%$ in Gr 8. We also investigated if 
the differences found between spiral and irregular galaxies in 
the [OIII]/H$\beta$ and the [NII]/H$\alpha$ ratio are due to differences in 
the metal content between these types of galaxies. Although the number 
of galaxies studied is not very large, it can be concluded that the [OIII]/H$\beta$ 
is not related with the oxygen content, while the situation is 
more ambiguous for the [NII]/H$\alpha$ ratio. 

{\it keywords galaxies: irregular --
galaxies: -- interstellar medium: H\,{\sc ii} regions:
general -- galaxies: individual: Gr 8 -- galaxies: individual: ESO 245-G05 }

\section{Introduction}

Extensive studies of the Diffuse Ionized Gas (hereafter DIG) have 
been carried out for spiral galaxies (e.g. Otte \& Dettmar 1999). These studies
show that   
the spectral line ratios for most of the 
spiral galaxies are very similar, with a very 
low [OIII]/H$\beta$, high values of [NII]/H$\alpha$, and [SII]/H$\alpha$  
both increasing with the distance to the plane,  and high values of the 
ionization, as defined 
by [OII]/[OIII] (T\"ullmann \& Dettmar 2000). On the other side,  the DIG 
properties
 of  irregular galaxies are not  known very well, as 
very few galaxies have been studied so far. These few data have  
prevented us to make any conclusion on their spectral characteristics (Martin 
1997; Hidalgo-G\'amez 2005; Hidalgo-G\'amez \& Peimbert 2006). These 
data show that the line ratios differ from those in spiral 
galaxies. The [OIII]/H$\beta$ is large. For some of the 
galaxies, e.g. IC 10, is larger than $1.5$. The [NII]/H$\alpha$ and 
[SII]/H$\alpha$ ratios 
are smaller than $0.5$. Moreover, if the line ratios are very different 
for spiral  and irregular galaxies, it is likely that the 
ionization source of the DIG will also differ 
due to different metal contents, gas masses or Star 
Formation Rates (SFR).  

In the present investigation we study two dwarf irregular galaxies with 
low chemical abundances and intermediate SFRs: Gr 8 and ESO 245-G05. The 
first one is a nearby galaxy at a distance of D = $2.24$ Mpc (Hidalgo-G\'amez 
\& Olofsson 1998) dominated in the optical by two large H\,{\sc ii} regions. 
The metallicity is very low (12+log(O/H) = $7.4 \pm 0.1$, Hidalgo-G\'amez 
\& Olofsson 2002) and 
also the SFR, which is $2.12 \times 10^{-3}$ M$_\odot$ yr$^{-1}$ (Hunter 
\& Elmegreen 
2004). ESO 245-G05 is one of the Sculptor group dwarf galaxies. The oxygen 
content is also low ($7.6$-$7.8$ with small variations along the bar, 
see Hidalgo-G\'amez et al.  2001) and the SFR is 
$1.1 \times 10^{-2}$ M$_\odot$ yr$^{-1}$ (Miller 1996). In both galaxies, 
a large extension of ionized gas in their H$\alpha$ images is present. Therefore, they 
represent good candidates for studying the properties of the DIG. Due to their 
low metallicities they could be used to  study  the dependence 
of the DIG properties with some of the characteristics of the 
irregular galaxies.

The paper is structured as follows: the observations and 
the reduction of the data are described in the next Section. In Section 3, the line ratios along 
the slit for each galaxy are presented.  Possible sources 
of ionization of the Diffuse Gas are described in Section 4. Section 5 
discusses the relation between the metallicity and the DIG properties. Conclusions are presented in Section 6. 

\section{Observations and data reduction}

The data were obtained with two different facilities. The spectrum of 
ESO 245-G05 was acquired on August 9th, 1997 with the 3.6m telescope at La 
Silla Observatory with EFOSC1. All the details about the data acquisition 
and the reduction procedure are described in Hidalgo-G\'amez et al. (2001) 
and, therefore, are not going to be repeated here. We will remember 
that the slit position used for this analysis had a position angle of 
$295$ degrees, covering the southern part of the bar (see Figure ~\ref{esoha}), and the 
real spectral resolution was $18$\AA. These data were  acquired and 
reduced without considering the low surface brightness emission and, 
therefore, the sky subtraction was severe in order to eliminate the 
strongest sky lines.   

The spectra of Gr 8 were obtained on March 13th-14th, 2002 with the 
2.1-m telescope of the Observatorio Astron\'omico Nacional at San Pedro 
M\'artir (OAN-SPM). The Boller \& Chivens spectrograph was used with a 
300 l/mm grating blazed at 5000 \AA. The slit width was 170\,$\mu$m, 
subtending $\approx$ 2${\arcsec}$ on the sky, and yielding a spectral 
resolution of $7$ \AA. The full spectral range observed was 3500-6900\AA. 
The first night (13/03/02) was not very photometric, due to cirrus and strong 
wind, whereas during the second night weather conditions were good, 
 although the seeing was high. Table 1 shows the log 
of observations. No correction for differential refraction was performed 
as the pixel size was small and also the airmasses were smaller than $1.3$ in all 
the 
slit positions but one. The orientation of 
the slits was E-W with a total of six slit positions.

The reduction of the Gr 8 data was performed with the MIDAS software 
package. Bias and sky twilight flatfields were used for the calibration of 
the CCD response. Due to the strong differences in the sensitivity of the 
CCD as to the wavelength, the spectral range was divided into three parts: 
from 5200 \AA~ to 6800 \AA~ (red), from 4000 \AA~ to 5600 \AA ~(green) and 
from 3500 \AA~ to 4800 \AA~ (blue). Unfortunately, due to the low efficiency 
of the spectrograph at the blue end, no spectral lines were detected there 
and, therefore, they were not used in the present investigation.  He-Ar lamps were 
used for the wavelength calibration.  The spectra were corrected for 
atmosferic extinction using the San Pedro M\'artir tables (Schuster \& 
Parrao 2001). Several standard stars were 
observed each night in order to perform the flux calibration. The accuracy 
of the calibration was better than 5$\%$ for both nights. The most 
difficult task was the sky subtraction because removing the 
strong sky lines of the spectra will also remove most of the 
low surface brightness emission. Therefore, only a dozen of rows 
were used in the sky templates for each spectrum and only the underlying sky structure was subtracted, but most of the stronger sky lines still remain 
in the spectra. Special care was taken in the analysis procedure for these 
data. However, it will not affect the results because a strong sky 
subtraction was perfomed on the  ESO 245-G05 data, being the results 
very similar (see Section 3).

Both sets of data, fully reduced and calibrated, were divided into a number 
of spectra, covering three rows for ESO 245-G05 and five for Gr 8 
(hereafter, the r-spectra). The total sizes are $1.8$ and $5$ arcsec, respectively. These values match the seeing conditions 
and corresponds to $39$ pc in ESO 245-G05 and $40$ pc in Gr 8. The total 
number of spectra studied was $51$  for the slit position of ESO 245-G05 
and $77$ for the six positions in Gr 8. 

Finally, we wish to comment on the uncertainties. Three different sources 
were considered: the uncertainties in the level of the stellar continuum 
with respect to the line,  $\sigma_c$, those introduced by the 
reduction procedure, $\sigma_r$, (especially flatfielding and flux 
calibrations) and uncertainties due to the extinction correction, 
$\sigma_e$.   The final uncertainty for each line was determined from 
$$\sigma = \sqrt {\sigma_c^2 + \sigma_r^2  + \sigma_e^2}$$ These 
uncertainties were measured for each line at each spectrum in all the 
positions for each galaxy. With these values a total uncertainty for 
each line and each slit position can be determined, and is presented in 
Table 2.

\section{Diffuse Ionized Gas}

As previously said, the main purpose of this investigation is to obtain 
the spectral characteristics of the Diffuse Gas in the mentioned irregular galaxies.
Spiral galaxies are the main targets in DIG studies, mainly because 
the DIG is situated above the plane of the disk (e.g. Otter \& Dettmar 
1999) or in the inter-arm region (Benvenuti et a. 1976). In the case of 
the Milky Way, the low density is the main characteristic for the discrimination.

The difficulties of studying DIG in irregular galaxies are due to the problem
of distiguishing between the ionized  
gas inside and outside the borders of the classical   
the H\,{\sc ii} regions. A disk is not clearly defined in these galaxies. Moreover, density 
is one of the most difficult parameter to determine from 
long-slit spectroscopy data. The only way to do so is by using  the ratio of 
the lines of any of the following doublest: 
[OII]$\lambda$$\lambda$3726,3729\AA, [SII]$\lambda$$\lambda$6717,6731\AA, 
[ClIII]$\lambda$$\lambda$5517,5537\AA~ and [ArIV]$\lambda$$\lambda$4711,4740\AA~
(Aller 1984). Therefore, any of 
these doublets should be detected and resolved with high accuracy. The one  mostly 
used is [SII]$\lambda$6717/[SII]$\lambda$6731 because the intensity of the 
lines is high enough to be detected, and not very high spectral resolution 
is needed. The main drawback is that in the low density regime the 
differences in the [SII]$\lambda$6717/[SII]$\lambda$6731 between the 
H\,{\sc ii} regions, with typical densities of $100$ cm$^{-3}$, and the DIG 
with values of $10$ cm$^{-3}$, are about $0.1$ (see figure 5.3 in Osterbrock 
1989), which normally is smaller than the uncertainties in the ratio. The 
situation is even  worse for the other doublets. 
Thus, the density cannot be used for the discrimination of the DIG and 
the H\,{\sc ii} regions. 

Another parameter involved in the definition of the DIG is the Emission 
Measure (EM)  related to surface brightness (SB) in H$\alpha$ 
and the electronic temperature (Greenawalt et al. 1998). In 
the Milky Way there is a correlation between low EM and low density. Therefore, the EM can be used in the discrimination between DIG and H\,{\sc ii} regions when the density cannot be determined.  The 
main difficulty is the lack of an unique value for the EM. It varies from $80$ pc cm$^{-6}$ for 
the arms in spiral galaxies (Hoopes \& Walterbos 2003) to $2$ pc cm$^{-6}$ 
for the Milky Way (Reynolds 1989). Instead of choosing a value in this range, 
we decided to use another approach. The cumulative distribution function of 
the surface brightness in H$\alpha$ is a smoothly increasing function  with 
a change in the slope when H\,{\sc ii} regions begin to dominate the 
light. This SB(H$\alpha$) turnoff point can be considered as the discriminator between DIG and H\,{\sc ii} regions. This procedure has already been used 
for the two previous galaxies studied by the author (IC 10, 
Hidalgo-G\'amez 2005; NGC 6822, Hidalgo-G\'amez \& Peimbert 2006). It has to be kept in mind that this procedure, as any statistical method, is more accurate for a large number of data points especially when are included both cores of the H\,{\sc ii} regions and very weak emission regions.

\begin{figure}
\centering
\includegraphics[width=8cm]{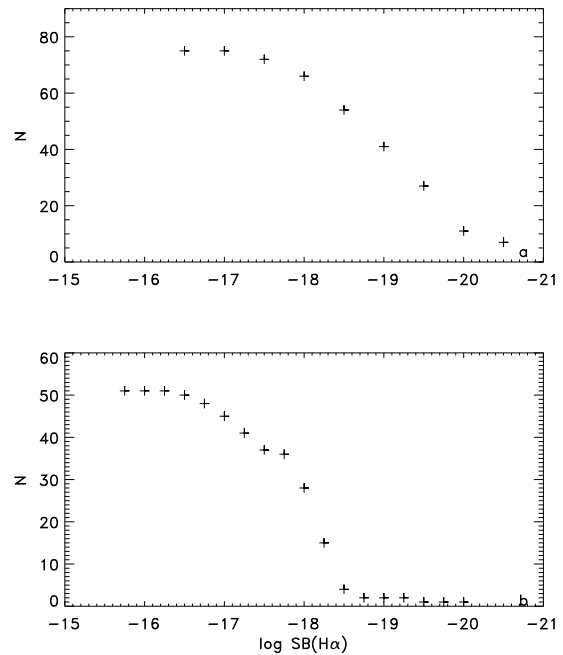}
\caption{The cumulative distribution function of the surface brightness in H$\alpha$, Galactic extinction corrected, for Gr 8 (panel a) and ESO 245-G05 (panel b). Both galaxies show a turn-off point at log SB(H$\alpha$) = -17.9, indicating the transition between DIG and H\,{\sc ii} regions.}
\label{sb}
\end{figure}

Figure ~\ref{sb} shows this function for Gr 8 and ESO 245-G05.  
Both galaxies have a turn-off point of log SB(H$\alpha$) $\approx$ -17.9, 
which corresponds to $1.3\times 10^{-18}$ erg cm$^{-2}$ s$^{-1}$
arcsec$^{-2}$. The surface brightnesses have been corrected for 
Galactic extinction following Schelgel et al. (1998). Another turn-off point 
is detected, especially in ESO 245-G05 at $2\times 10^{-19}$ erg 
cm$^{-2}$ s$^{-1}$ arcsec$^{-2}$, which might be an indication of the 
noise level. There are only two data points with SB(H$\alpha$) lower than this 
limit and they are not going to be considered hereafter.  
 Therefore, we will regard as DIG data those spectra with 
SB(H$\alpha$) between $2$ and  $13~ 10^{-19}$ erg cm$^{-2}$ 
s$^{-1}$ arcsec$^{-2}$. The latter value is slightly lower than in the other 
two irregular galaxies studied by the author, which gives log SB(H$\alpha$) = 
$-17.45$ for 
NGC 6822 (Hidalgo-G\'amez \& Peimbert 2006) and log SB(H$\alpha$) = $-17.5$ 
for IC 10 
(Hidalgo-G\'amez 2005), both galaxies  having larger SFRs and metallicities. This 
range splits the spectra into $21$ H\,{\sc ii} regions and $56$ DIG 
locations in Gr 8. The corresponding values in ESO 245-G05 are $20$ and 
$29$, respectively. In addition, two  points 
with  surface brightness on the boundary between the DIG 
and the H\,{\sc ii} locations were found in this last galaxy.  

\begin{figure}
\centering
\includegraphics[width=8cm]{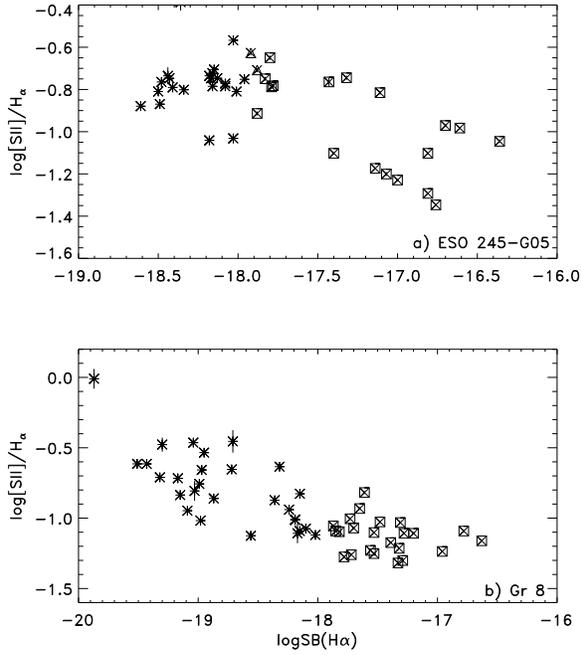}
\caption{The log SB(H$\alpha$) vs. [SII]/H$\alpha$ for Gr 8 (panel b) and ESO 245-G05 (panel a). DIG locations are plotted as stars and H\,{\sc ii} as squares. The two transition points in ESO 245-G05 are plotted as triangles (see text for a definition).  The error bars associated with every data fromt he r-spectra are presented. The data points shows an anticorrelation for Gr 8 while the DIG data points exhibit a  scatter diagram for ESO 245-G05.} 
\label{has}
\end{figure}

\begin{figure}
\centering
\includegraphics[width=8cm]{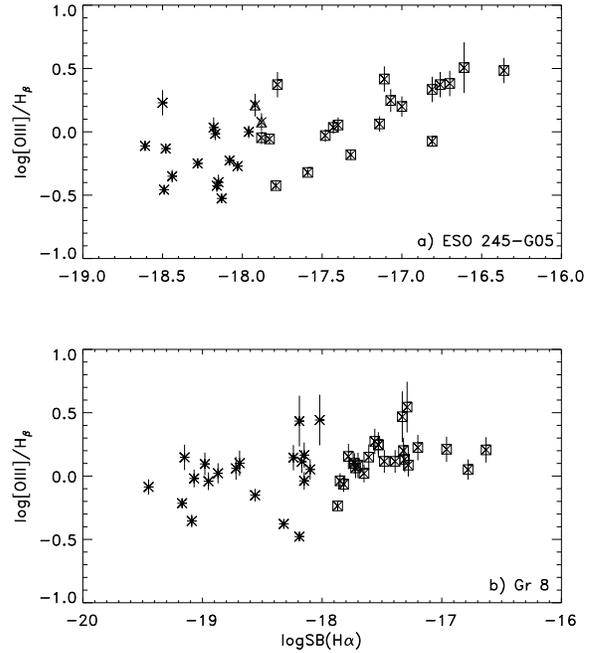}
\caption{The log SB(H$\alpha$) vs. [OIII]/H$\beta$  for Gr 8 (panel b) and ESO 245-G05 (panel a). Symbols as in Figure 4.  There is a clear correlation for the H\,{\sc ii} regions of ESO 245-G05 that become flatter for the DIG locations. The correlation is less step for Gr 8 with no abrupt change in the slope between DIG and H\,{\sc ii} regions. }  
\label{hao}
\end{figure}

In order to  test the correctness of this cumulative function for 
defining H\,{\sc ii} regions we can study the relation between 
the SB(H$\alpha$) and the line ratios. Fig ~\ref{has} shows the 
log SB(H$\alpha$) vs. [SII]/H$\alpha$ for ESO 245-G05 (a) and Gr 8 (b) for all the data points of the r-spectra. There
is a trend toward having lower [SII]/H$\alpha$ for larger SB, especially for Gr 8 while there is a turn-off point at log SB(H$\alpha$) = - 18.1 for 
ESO 245-G05. There is also  a bifurcation for the H\,{\sc ii} regions which could be related with diferences in the chemical content (Martin 1997). An opposite trend is shown in Figure ~\ref{hao} between log SB(H$\alpha$) 
vs. log [OIII]/H$\beta$ for both galaxies. The smooth transition between DIG locations and 
H\,{\sc ii} regions indicates that the distinction between them is 
correct. Another way to verify it is to compare 
the size of the H\,{\sc ii} regions determined from the cumulative function 
and the extension of the spectral lines. The sizes obtained with 
the cumulative function correspond to the high ionization regions. Therefore, although the cumulative function 
is a statistical approach, it serves as a good distinction between DIG 
and H\,{\sc ii} regions when density cannot be determined.

In addition to the r-spectra defined in the previous section, 
the so-called integrated spectra were obtained summing up 
all the pixels with surface brightness lower than 
$1.3\times 10^{-18}$ erg cm$^{-2}$ s$^{-1}$ arcsec$^{-2}$ (DIG) and higher 
than it (H\,{\sc ii}) to obtain spectra of high S/N for each 
H\,{\sc ii} and DIG region.   

In the following, we will study the variations along the slits of the 
line ratios of interest for each galaxy. 

\subsection {ESO 245-G05}

As already said, there is only one slit position for this galaxy. In both 
the r-spectra and the integrated ones, the recombination lines H$\alpha$, H$\beta$, 
as well as the forbidden ones  [OII]$\lambda$3727\AA, [OIII]$\lambda$5007\AA, 
[NII]$\lambda$6583\AA, [SII]$\lambda$6717\AA~ 
and [SII]$\lambda$6731\AA, were determined. The hidrogen line H$\gamma$ 
was detected at very few locations, 
mainly corresponding to H\,{\sc ii} regions. The intensities of all the 
lines were normalized to H$\beta$, and the ratios [OII]$\lambda3727$/H$\beta$, 
[OIII]$\lambda$5007/H$\beta$, [NII]$\lambda$6583/H$\alpha$, 
[SII]$\lambda$6717/H$\alpha$ and [SII]$\lambda$6717/[SII]/$\lambda$6731 
were obtained. The reddening correction is very important when 
comparing intensities of lines with very different wavelengths. 
Unfortunately, the ratio H$\alpha$/H$\beta$ cannot be used for the 
extinction determination, as explained in Hidalgo-G\'amez et al. (2001). In 
that 
investigation the H$\gamma$/H$\beta$ was used for the reddening correction. 
But, as previously said, the H$\gamma$ line was not detected elsewhere 
and therefore, it was not a solution here. In consequence, no 
reddening correction was performed. For the majority of the line 
ratios 
the wavelength range is 
very narrow and the values remain, essentially, unaffected. The only 
exception 
is the [OII]/H$\beta$ ratio where the differences 
between the non-corrected and the corrected values 
might reach up to  30$\%$. Therefore, we are not going to discuss this ratio 
here. Table 3 shows the line ratios  from the integrated spectra 
with a total of $3$ H\,{\sc ii} regions and $4$ DIG regions. Due to the 
smaller width of the slit and the high resolution of this configuration, 
the ratio [SII]$\lambda$6717/[SII]$\lambda$6731 is quite reliable.  The 
number of pixels of region {\it d1} is very small and so is 
the S/N. 
Therefore, it is not useful for obtaining any  conclusion. 

\begin{figure}
\centering
\includegraphics[width=8cm]{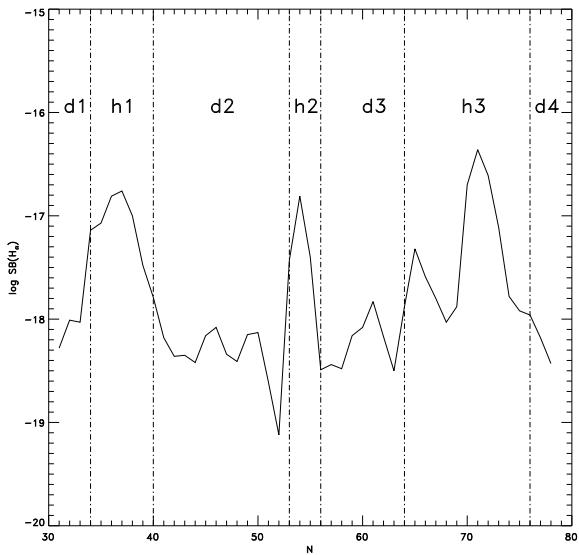}
\caption{The surface brightness along the slit position in ESO 245-G05. The H\,{\sc ii} regions, marked with their names, are located inside the dashed-dotted lines. Spectrum nr 52 is one of the data points which value is below the noise level.} 
\label{rendijasb}
\end{figure}

\begin{figure}
\centering
\includegraphics[width=8cm]{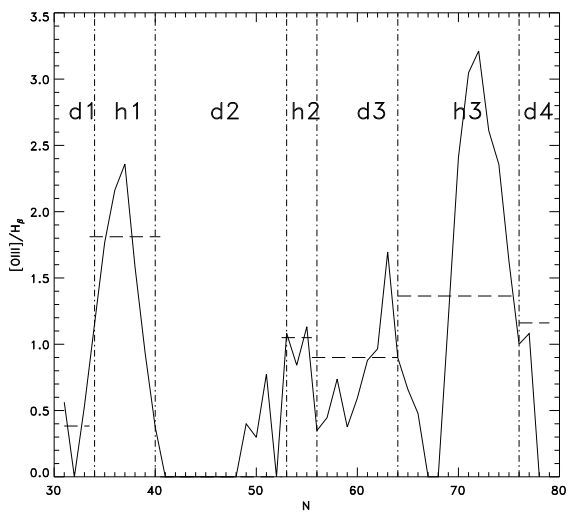}
\caption{The [OIII]/H$\beta$ along the slit position in ESO 245-G05. The H\,{\sc ii} regions are located inside the dashed-dotted lines. The long-dashed 
lines correspond to the values from the integrated spectra.  The largest excitation corresponds to the second condensation of {\it h3}}. 
\label{eoii}
\end{figure}

\begin{figure}
\centering
\includegraphics[width=8cm]{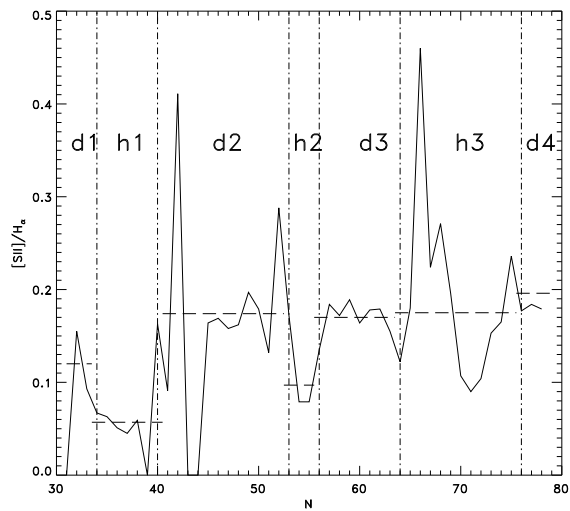}
\caption{The [SII]/H$\alpha$ ratio along the slit position in 
ESO 245-G05. Symbols as in Fig.~\ref{eoii}. It is very remarkable 
the large value of this ratio inside {\it h3} (see text for details).}
\label{es}
\end{figure}

\begin{figure}
\centering
\includegraphics[width=8cm]{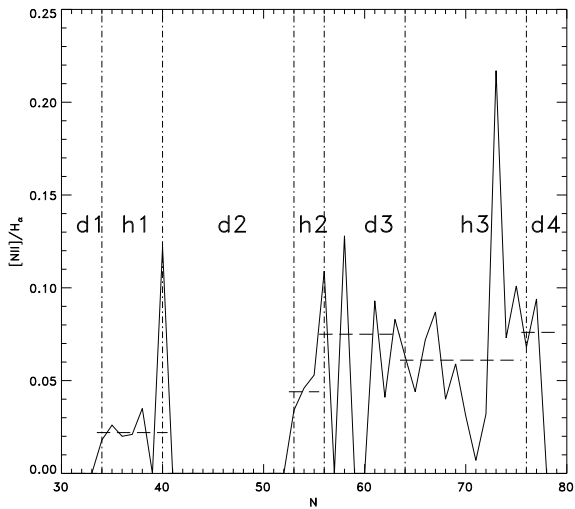}
\caption{The [NII]/H$\alpha$ ratio along the slit in ESO 245-G05. Symbols as in Fig. ~\ref{eoii}. Again, {\it h3} has the largest value of this ratio. On the contrary, nitrogen is no detected in any place of {\it d2} but one.} 
\label{en}
\end{figure}

The line ratios along the slit for the surface brightness, the excitation 
the [SII]/H$\alpha$ and the density are shown in figures from
~\ref{rendijasb} to ~\ref{en}.  
The H\,{\sc ii} 
regions are marked by dashed-dotted lines while the values from the 
integrated
spectra are shown by dashed lines. Fig.~\ref{rendijasb} shows the 
SB(H$\alpha$) along the slit. The most interesting feature is that {\it h3} 
is formed by two  condensations, with a region of low surface brightness between 
them. As there was only one data-point, it was considered  a 
transition point and part of the H\,{\sc ii} region. Moreover, the 
second condensation has the large SB in the region under study.   

The excitation, defined as [OIII]$\lambda$5007/H$\beta$, is larger 
inside H\,{\sc ii} regions (Fig. ~\ref{eoii}), being the only exception 
the spectra No {\it 61}-{\it 63}, where the [OIII]/H$\beta$ is larger than 
in {\it h2}. This is not considered  anomalous as these fluctuations are 
observed in other galaxies too (e.g. IC 10, Hidalgo-G\'amez 2005). It is 
very interesting that the 
[OIII]$\lambda$5007\AA~ line was not detected either the majority of  
{\it d2} spectra or in the integrated spectra, which is an indication of very 
few [OIII]/H$\beta$ of the DIG in this region. Other interesting feature is the 
very low [OIII]/H$\beta$, dropping to 0 at the transition point in the eastern condensation of 
{\it h3}. 

The second interesting line ratio is [SII]$\lambda$6717/H$\alpha$ which is 
shown in figure ~\ref{es}. From Table 2 it could be seen that the values of this ratio 
increase along the slit towards the west for H\,{\sc ii}, while it is 
almost constant for DIG locations. The very large 
values of this ratio inside the H\,{\sc ii} regions are very disturbing. Such values are very 
unusual as normally, this ratio is  very low inside H\,{\sc ii} regions, 
lower than $0.05$ (Hidalgo-G\'amez \& Olofsson 2002), but it reaches up 
to $0.47$ inside {\it h3}. This high value coincides with the first 
condensation of this region while the transition point has a value 
of $0.27$. In the case of the other two H\,{\sc ii} regions, the values 
(both integrated and along the slit) are very similar to what is obtained 
for  other irregular galaxies (e.g. Hidalgo-G\'amez \& Peimbert 2006). 
 As far as  the DIG locations concern, there are two peaks of $0.41$ and $0.29$ at 
both ends of {\it d2}, but the averaged values are very similar to those in 
Gr 8 (see Section 3.2). As it is well known, 
for values larger than $0.3$, shocks might be important contributors to 
the ionization of the lines 
(Dopita 1993). Except for these three spectra 
(No {\it 41}, {\it 52} and {\it 66}), the values indicate that shocks 
are not  important elsewhere. 

Finally, we can discuss the ratio [SII]$\lambda$6717/[SII]$\lambda$6731, 
which is related with the density (Osterbrock 1989). DIG locations have 
smaller density and, therefore, larger [SII]/[SII] ratios. From Table 3 we 
see that all the regions, except maybe {\it h1} and {\it d2}, are inside the 
low density limit. The exact value of the density cannot be determined, 
as previously discussed. Moreover, some of the values are above the 
theoretical limits due to the blending of the lines, which is an indication 
of very low densities.
  
\subsection{Gr 8}

A similar procedure was used for Gr 8. Because 
of    the low efficiency of the B\&Ch spectrograph at the blue 
end, [OII]$\lambda$3727\AA~was not detected. The  lines H$\alpha$, 
H$\beta$, [OIII]$\lambda$5007\AA, 
[NII]$\lambda$6583\AA, [SII]$\lambda$,$\lambda$6717,6731\AA~ were 
measured in the r-spectra.  Due to the spectral resolution of this configuration, the two lines of [SII] cannot be resolved properly and we prefered to have only a measured of both lines together. Their intensities were normalized to H$\beta$ and 
extinction corrected using the H$\alpha$/H$\beta$ ratio. Because the 
line [NII]$\lambda$6583\AA~ was detected in only a few locations along each 
slit position, it is not interesting to study its variations along the 
slits. In fact, it was detected in only one DIG integrated spectrum. 
Thus it was not a configuration problem but a real deficiency of nitrogen in 
this galaxy.  Moreover, Fabry-Perot interferometry indicates that nitrogen 
is located at a very small part of the western H\,{\sc ii} region 
(M. Rosado, private communication).
In addition, the recombination line He I $\lambda$5876\AA~ was detected in 
four  H\,{\sc ii} regions of the integrated spectra. Therefore, the ratios 
of interest in Gr 8 are only [OIII]$\lambda$5007/H$\beta$ 
and [SII]$\lambda\lambda$6716,6730\AA/H$\alpha$. Besides these, the 
surface brightness and the extinction along each slit can also be studied. 

In Table 4 the surface brightness, the [OIII]/H$\beta$ and the 
[SII]/H$\alpha$ ratios are  given for all the integrated spectra, as well as 
the extinction coefficient. The values of [NII]$\lambda$6584\AA~ 
and HeI$\lambda$5876\AA~ are also tabulated when detected. 

\begin{figure}
\centering
\includegraphics[width=6cm]{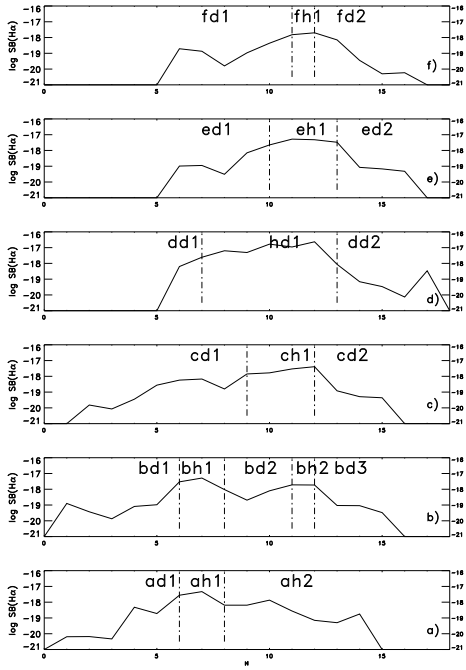}
\caption{Plot of the SB(H$\alpha$) 
along the slits in Gr 8. Slit {\it f} corresponds to the northen part of the galaxy and slit {\it a} to the southern. There is a migration of the H\,{\sc ii} regions from the west (left in the figure) to the east of the galaxy. }
\label{sbslit}
\end{figure}

Figure ~\ref{sbslit} shows the surface brightness in H$\alpha$ along the 
slits for each of the six slit positions. The x-axis corresponds to the number 
of the spectra. The orientation is from the north of the galaxy in the upper part 
of the figure (slit f) to the south at the bottom (slit a).  
The H\,{\sc ii} regions are marked between  dash-dotted lines. As seen in 
Table 1, the distance 
between the slit positions is not the same for all of them. Therefore the H\,{\sc ii} regions are not coincident. In any case, 
it is interesting to notice the movement of the emission from west (at the 
left) to east (at the right). Also, the H\,{\sc ii} regions move in the 
same direction. This displacement is not due to the movement of the 
galaxy on the sky which would give a displacement in the opposite 
direction but to the orientation of the galaxy, which is south-west to north-east, as observed in this Figure.  Finally, the sizes of the H\,{\sc ii} regions vary from only $37$ 
pc to $200$ pc, fitting the orientation of the galaxy. With this information we can identify the H\,{\sc ii} regions with the most prominent regions named nr 5 in slits {\it a)} and {\it b)} and nr 19 in the rest of the positions of Hodge et al. (1989). 

\begin{figure}
\centering
\includegraphics[width=6cm]{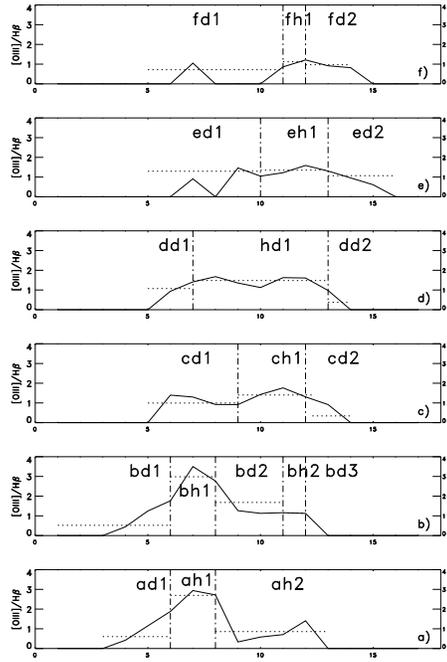}
\caption{The [OIII]/H$\beta$ along the slits in Gr 8. In addition to the lines in Fig.~\ref{sbslit}, the dotted lines correspond to the value of the integrated spectra. The largest [OIII]/H$\beta$ is located at the south part of the galaxy (region No 5 in Hodge et al. 1989), decreasing towards the north. The migration of the [OIII]/H$\beta$ to the east is also very prominent.  }
\label{oslit}
\end{figure}

Figure ~\ref{oslit} shows the [OIII]/H$\beta$. The orientation is the same as 
in Figure ~\ref{sbslit}. In addition to the H\,{\sc ii} region positions, 
the values of the integrated spectra as shown in Table 4 are given 
by the dotted lines. All the slit positions have the same range in 
[OIII]/H$\beta$ in order to see their differences throughtout the galaxy. 
In this sense, slits (a) and (b) have the largest [OIII]/H$\beta$ ratio, while 
it decreases towards the north, reaching a minimun at slit (f). Considering 
the [OIII]/H$\beta$ throughout each slit position,  the H\,{\sc ii} 
regions have, in general, the largest [OIII]/H$\beta$, but there are 
exceptions. In fact, there is not a general guideline. For example, the 
values of the [OIII]/H$\beta$ are very different in H\,{\sc ii} and 
DIG locations for slit {\it (a)}, 
while they are of the same order in slits {\it (e)} and {\it (f)}. 
In the case of slits {\it (c)} and {\it (d)} the [OIII]/H$\beta$ of the 
DIG is larger towards the west than towards the east. This agrees with the surface brightness distribution of Hodge et al. (1989). For all the 
positions, there is a good agreement between the values for the 
integrated spectra and the r-spectra. Indeed, the largest values of 
the integrated spectra correspond to {\it bh1} and {\it ah}, which are also 
the locations with the largest [OIII]/H$\beta$s in the r-spectra.  
Another important fact to be noticed is that, contrary to the 
previous irregular galaxies study by the author (IC 10 and NGC 6822), there 
are DIG locations with very low 
[OIII]/H$\beta$, lower than $0.5$, such as {\it cd2} and {\it dd2}. 

\begin{figure}
\centering
\includegraphics[width=6cm]{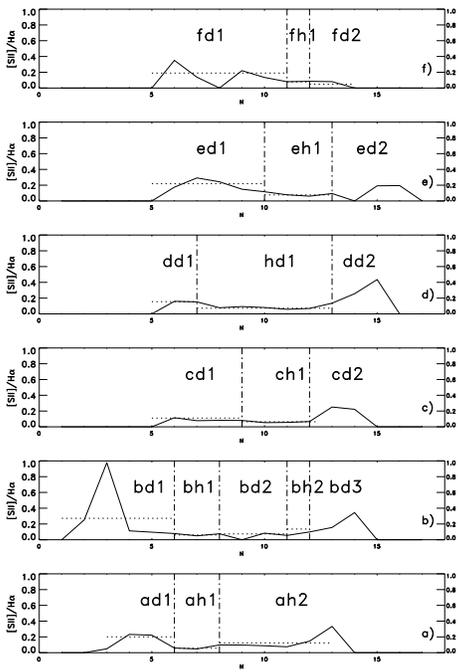}
\caption{The [SII]/H$\alpha$ ratio along the slits in Gr 8. Symbols as in Fig. ~\ref{oslit}. Except for the spectrum nr 3 in {\it bd1} and nr 15 in {\it dd2}, the rest of the values are smaller than $0.3$, indicating the absence of shocks.}
\label{sslit}
\end{figure}

Figure ~\ref{sslit} shows the [SII]/H$\alpha$ ratio. The orientation and the meaning of the lines  as in Fig. ~\ref{sbslit}. The range in all 
the slits is the same, between $0$ and $1$. The highest value of this ratio is 
in slit {\it b)}, with [SII]/H$\alpha$ = $1$, but only in 
slits {\it f} and {\it d} some locations have values larger than $0.3$.  The value of 
spectrum {\it bd1} is very striking because although the H$\alpha$ images of 
this galaxy reveal two large H\,{\sc ii} regions (Hodge 1967) with star 
forming activity, the Star Formation Rate is low, only 
$2.2\times10^{-3}$ M$_{\odot}$ yr$^{-1}$ (Hunter \& Elmegreen 2004). Moreover, there are 
no superbubbles reported in this galaxy which might explain the large values 
of the [SII]/H$\alpha$ ratio. Begum \& Chengalur (2003) interpreted 
their results in H\,{\sc i} for this galaxy as a combination of rotation and 
a radial movement (expansion outwards because of the star formation activity 
or infall inwards due to the formation process which has not been completed 
yet). But in their maps, there is no particular disturbance  at the locations 
of spectra {\it bd1} or {\it dd2}. In any case, 
the results of Fig. ~\ref{sslit} are in agreement with those from 
Fabry-Perot interferometry of this galaxy (Rosado, private communication). 
In general, the values of [SII]/H$\alpha$ are smaller inside H\,{\sc ii} 
regions than in DIG locations. However, there are a few places where 
these differences are very small, as occurs in {\it bd2}. Once again, the agreement 
between the integrated and the r-spectra is very good, in general. 

\begin{figure}
\centering
\includegraphics[width=6cm]{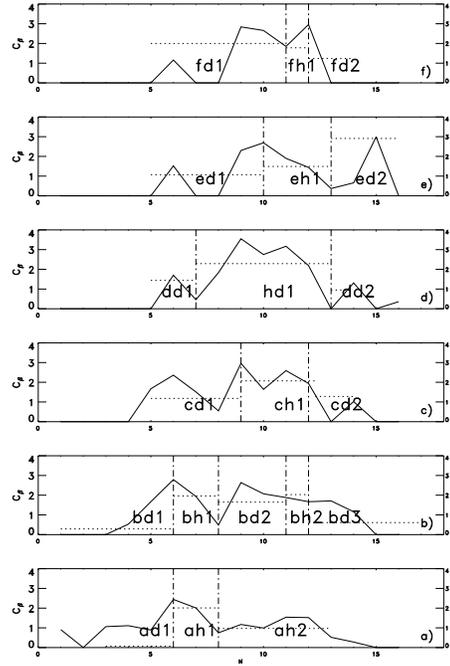}
\caption{The extinction coefficient along the slits in Gr 8. Symbols as in Fig. ~\ref{oslit}. We see that the extinction is not homogeneous inside the galaxy, as concluded from the studies in H\,{\sc i}.  The Galactic extinction towards this galaxy corresponds to a C$_{\beta}$ = $0.04$.}
\label{cb}
\end{figure}

Begum \& Chengalur (2003) also noticed that the H\,{\sc i} is clumpy, and they considered it is associated with the optical emission. A  similar conclusion can be reached 
from Figure ~\ref{cb},  showing the extinction coefficient along the 
slits. From this figure, a very interesting result can be inferred. 
The concentration of dust is not homogenous along the slits and  does 
not follow any pattern. The largest  C$_{\beta}$ values of some of the slits 
are inside the H\,{\sc ii} regions (slit {\it (d)}) while in other cases, 
such as slit {\it (e)}, the maximun value of the C$_{\beta}$ is found  at a DIG 
location. 
Apparently, these results do not agree with those from Begum 
\& Chengalur (2003). A possible explanation is that their large observing resolution 
prevent them from observing the interior of the 
H\,{\sc ii} regions, as we have done here. In fact, there is a good agreement 
between both investigations, as  the extinction values are very large in 
all but one position ({\it bd1}).

The spectral characteristics of the emission lines in ESO 245-G05 are 
very different from what is observed in other galaxies, not only at the 
DIG locations but also at the H\,{\sc ii} regions, due to the very large values 
of [NII]/H$\alpha$ and [SII]/H$\alpha$.  The situation in Gr 8 
is more similar to what is found in other galaxies. 
In the following, we are going to find the model which  fits best all the 
line ratios in both galaxies.

\section{The ionization source}
 
\subsection{Radiation bounded vs. density bounded H\,{\sc ii} regions}

The theory of  H\,{\sc ii} regions  was developed by Str\"omgren in 1939 
and since then they have been considered as radiation bounded (Aller 
1984). In this case, there is not enough nearby neutral gas to absorb the total Lyman continuum of the source and form a classical Str\"omgrem sphere.  With 
such a model the line ratios can be predicted when the ionizing temperature 
of the OB associations is known (Osterbrock 1989). This is true not only for the region inside 
the Str\"omgren radius (the classical H\,{\sc ii} region) but also for the 
gas outside this limit, the Diffuse Ionized Gas. Mathis (1986) and 
Domg\"orgen \& Mathis (1994) made preditions on the line ratios of the DIG 
for different ionization parameters. They concluded that both 
[NII]/H$\alpha$ and [SII]/H$\alpha$ are larger than $0.3$ and the 
[OIII]/H$\beta$ is very low ($<$ $0.1$). Another important result is 
the constancy of the [NII]/H$\alpha$ ratio between the DIG and the 
H\,{\sc ii} regions (Mathis 1986), as well as a lack of correlation 
between this ratio 
and [SII]/H$\alpha$. In recent years it has been considered that 
H\,{\sc ii} regions may be density instead of radiation bounded (Beckman et al. 2000). It would means 
that photons still would have enough energy to ionize the gas outside the 
classical Str\"omgren sphere.  Moreover, considering inhomogeneities in 
the density inside the H\,{\sc ii} region (Giammanco et al. 2004), the 
escape of photons is not the same in all  directions. Using such a model,  
Hoppes \& Walterboos (2003) predicted the line ratios of the DIG for 
an Orion metallicity and an ionization parameter of $10^{_3}$. 

We can compare the data in Tables 3 and 4 with the predictions of both 
models by taking into account that they both were made for higher 
metallicity regions than those considered here. A lower metallicity implies 
a high ionization parameter, which implies higher [OIII]/H$\beta$. Moreover, 
a consequence of the leakage of photons is a harder spectrum, therefore
 higher  [OIII]/H$\beta$ for the same T$_{ion}$. 

Classical photoionization theory (radiation bounded) cannot explain the 
line ratios of these galaxies. The main reasons are that even though 
the [OIII]/H$\beta$ is lower in these two galaxies than in other 
irregular galaxies, it is not as low as $0.1$ in any place, and only lower 
than $0.4$ in {\it cd2} and {\it dd2} in Gr 8 and in {\it d1} in 
ESO 245-G05. Also, neither [NII]/H$\alpha$ nor [SII]/H$\alpha$ show very 
large values but only in those spectra discussed in sect. 3. There is 
no improvement either when lower metallicities are considered (see Hidalgo-G\'amez 
\& Peimbert 2005 and references therein). 

\begin{figure}
\centering
\includegraphics[width=8cm]{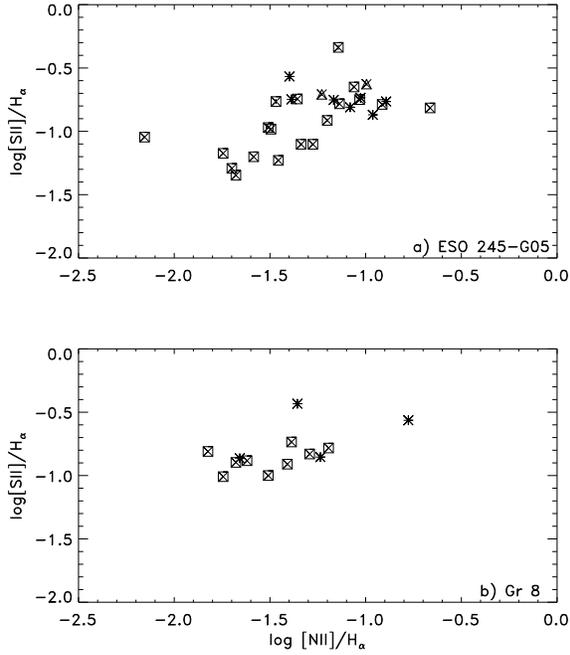}
\caption{The diagnostic diagram log[SII]/H$\alpha$ vs. log[NII]/H$\alpha$ for ESO 245-G05 (panel a) and Gr 8 (panel b). Stars correspond to DIG locations and squares to H\,{\sc ii} regions.  All the data points from the r-spectra with their uncertainty are presented in this plot. A strong correlation is observed between these two parameters at the DIG in spiral galaxies. Such a correlation is not presented here but a trend.}
\label{logsn}
\end{figure}

Another argument is found in Figure  ~\ref{logsn} which shows the log~[NII]/H$\alpha$ 
vs. log~[SII]/H$\alpha$ for ESO 245-G05 and Gr 8. The symbols are the same 
for both galaxies: diamonds stand for H\,{\sc ii} regions, stars for DIG and, 
in the case of ESO 245-G05, the two transition points are represented by 
triangles. These two parameters show a correlation for the DIG in 
spiral galaxies which cannot be explained by photoionization (e.g. Rand 
1998). Recently, Wood \& Mathis (2004) reproduced such correlation 
using Montecarlo simulations with photoionization models. They explained this 
correlation by the temperature increase  away from the ionizing 
sources with the increasing distance from the plane. There is 
a correlation in Figure ~\ref{logsn} between H\,{\sc ii} regions of 
ESO 245-G05, while a scatter diagram is given for DIG locations of this 
galaxy and for all the data points of Gr 8. This figure cannot be explained 
by any change in the Wood \& Mathis's models. In irregular galaxies 
the ionizing sources are located in all directions and therefore there are 
no important changes in the temperature. Another explanation might be variations in the abundances of nitrogen or sulfur. Irregular galaxies are considered as chemicaly homogeneous galaxies and only NGC 5253 shows a local enhancement of nitrogen, probably due to the WR population (Kobulnicky et al. 1997). Nothing can be said about the nitrogen distribution in ESO 245-G05 because it was measured in only one H\,{\sc ii} region. Considering Gr 8, Moles et al. (1990) reported differences in the log N/O of $0.28$ while only $0.03$ in log O/H between two H\,{\sc ii} regions. If these are real nitrogen variations they can explain the lack of correlation observed in Figure ~\ref{logsn}.

We are going to check if photon leakage models (density bounded regions) 
can explain better the intensities of the line ratios observed. The study of ESO 245-G05 
is easier because there are only four DIG regions 
(see Table 2). For all of them, both the T$_{ion}$ and the percentage of photon 
leakage agree when all the line ratios are considered. Values of  $40 \%$ 
from the [OIII]/H$\beta$, $30 \%$ from the [NII]/H$\alpha$, and between 
$35$ to $80 \%$ from [SII]/H$\alpha$ are obtained. The T$_{ion}$ is between $35,000$ K 
and $40,000$ K in all the cases. Considering the low metallicity of this 
galaxy, a leakage of photons of $35 \%$ and an ionization temperature 
of $40,000$ K can account for all the three line ratios.   

The study of the DIG in Gr 8 is more difficult, not only because of the large 
number of locations but also due to the wide range in temperatures 
and leakages. Moreover, the nitrogen ratio was detected only in one DIG 
location and therefore only the [OIII]/H$\beta$ and the [SII]/H$\alpha$ 
are used. Table 5 shows the percentage of leakage and the T$_{ion}$ from 
the two ratios. To explain the [OIII]/H$\beta,$ a leakage of at least 
$50 \%$ is needed, while the [SII]/H$\alpha$ gives values no larger than 
$40 \%$. Surprisingly, both ratios give similar temperatures of 
$\approx$ $35,000$-$40,000$ K. Considering that the metallicity of this 
galaxy is extremely low, 12+log(O/H) = $7.4 \pm 0.1$ (Hidalgo-G\'amez 
\& Olofsson 2002),  metallicity can be regarded as the reason for
the high values of the [OIII]/H$\beta$.  

A quick inspection of Table 5 shows that the leakage needed from 
the [OIII]/H$\beta$ if lower abundances are considered, is still much 
larger than the leakage from the sulfur ratio ($\approx~ 35$ \%). 
One explanation could be that the emission from these two lines  
originates in two different regions. In order to produce 
[OIII]$\lambda$5007\AA, very energetic photons are needed. Therefore, 
the regions where this line originates must be closer to the H\,{\sc ii} 
regions with T$_{ion}$ larger than  $40,000$ K, while the 
[SII]$\lambda$6717\AA~ might be originated futher away from the 
H\,{\sc ii} regions and, therefore, the temperature is lower and 
the photons softer. Moreover, as we said before, the slits covered a large 
part of the galaxy and therefore,  conditions can vary for each 
slit position.  For example, slits {\it (a)}, {\it (b)} and 
{\it (f)} might have a large leakage, of $50 \%$ or even larger for 
slit {\it (b)}, in the [OIII] region. The slits  {\it (c)} and {\it (d)} 
are the most problematic to explain because  the only values of T$_{ion}$ and
leakage from 
[OIII]/H$\beta$ are very different from the range obtained from the 
[SII]/H$\alpha$ ratio. 

Therefore, in order to explain the ionization of the Diffuse Gas in Gr 8, 
not a single value of the leakage of photons but several of them should be used, 
because  physical conditions (density, ionization temperature, etc) 
vary along the galaxy.

\subsection{An extra ionizing source from shocks or turbulence?}

We can use the models from Dopita \& Sutherland (1995) to study 
the contribution of shocks to the ionization of the DIG. From their table 1 
it can be obtained that the [SII]/H$\alpha$ ratio is larger than $0.18$ for 
a model of shock+precursor with a shock velocity of $200$ km~s$^{-1}$. 
Regions {\it d2}, {\it d3} and {\it h3} in ESO 245-G05 marginally reach 
this value whereas it is higher for regions {\it ad1}, {\it bd1}, {\it ed1} 
and {\it fd1} in Gr 8. The main objection is that the [OIII]/H$\beta$ is very 
high ($\approx 1.6$) when this shock velocity is considered and none of 
these regions have such values. Three of them have [OIII]/H$\beta$ smaller 
than $1$. 

\begin{figure}
\centering
\includegraphics[width=8cm]{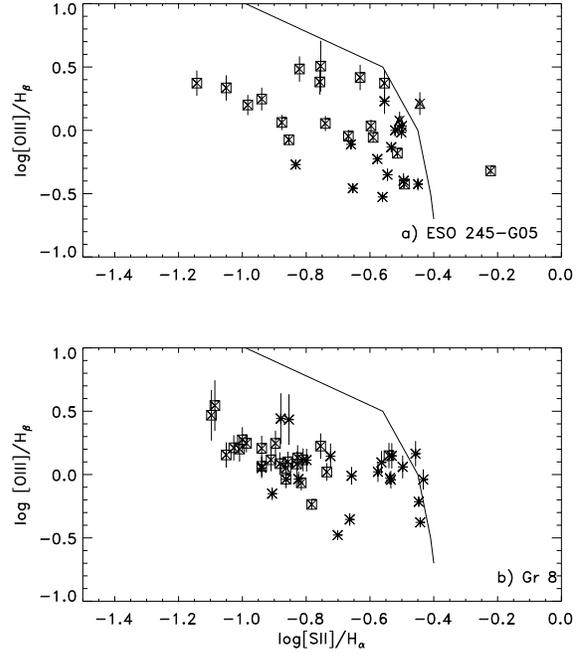}
\caption{The diagnostic diagram log[OIII]/H$\beta$ vs. log[SII]/H$\alpha$. 
Symbols as in figure ~\ref{logsn}. The line divides the diagram into the 
photoionized 
region (to the left) and the shocked one (to the right). Except for one H\,{\sc ii} of ESO 245-G05, the rest of the locations lie in the photoionized region.  }
\label{logso}
\end{figure}

Another important tool in the study of shocks are the diagnostic 
diagrams. Figure ~\ref{logso} shows the log[OIII]/H$\beta$ vs. log[SII]/H$\alpha$ diagram for both galaxies. The solid line divides the 
diagram into the photoionization (to the left) 
and the shocks (to the right) regions from Veilleux \& Osterbrock (1987). 
Only two  data points lie in the shocks region for ESO 245-G05, and none of 
them are DIG locations. In contrast, a few of DIG spectra in Gr 8 lie 
very close to the border line. These results are in agreement with those obtained in the previous paragraph, where very few locations might reach 
the values in the [SII]/H$\alpha$ ratio that would indicate shocks. In both cases we might expect that shocks play a more important role in these 
galaxies. The lack of detectable shock waves is interesting because the 
velocity field of ESO 245-G05 is very peculiar maybe because of 
past interactions (Cote et al. 2000).  A possible explanation is that only one slit position at the southern end of the galaxy is studied here. Gr 8 is kinematically supported by 
random motions (Begum \& Chengalur 2003)  and a total of 6 slit position, covering the most of the optical body of the galaxy have been studied. Therefore, it can be concluded that the gas motions inside Gr 8 are not very 
strong and do not have important consequences in the ISM.

Another interesting feature is the trend observed for ESO 245-G05, where 
the higher the [OIII]/H$\beta$, the lower the [SII]/H$\alpha$ is, with the 
DIG locations at the lower  end of the trend. On the contrary, the 
segregation in Gr 8 is more horizontal, with the DIG locations 
generally  closer to the shock line than the H\,{\sc ii} regions.

Finally, we can discuss another interesting model related to shocks as is 
the so-called Turbulent Mixing Layer by Slavin et al. 
(1993). This model contemplates the existence of a turbulent layer as 
the ionization source of the DIG. Slavin et al. predicted the line ratios with 
the transverse velocity of the hot gas (v$_t$), the intermediate 
temperature after mixing (T$^-$) and the abundance. When comparing here the 
values in Tables 3 and 4 with their figure 5 it is clear that 
[NII]/H$\alpha$ cannot be fitted with any set of parameters, whereas 
the [OIII]/H$\beta$ and [SII]/H$\alpha$ require differences in the 
T$^-$ of a factor of $2$, from $10^5$ to $2\times10^5$ K. This last result 
is not surprising as this model is optimized for low ionization ions 
(Slavin et al. 1993)

From this discussion we therefore conclude  that  no clear 
indication exists about  an extra ionization source of the DIG 
in these galaxies.

\section{Discussion} 

When comparing the DIG line ratios of spiral galaxies (Otte \& Dettmar 
1999) with those in irregular galaxies there appear two very striking 
characteristics: the higher [OIII]/H$\beta$ and the lower [NII]/H$\alpha$ 
ratio in irregulars compared to spirals. It is possible that the main reason for such a 
difference is the different metal content between these two type of 
galaxies. Irregular galaxies have metallicities in the range of $1/15$-$1/30$ Z$_{\odot}$ (e.g. Hidalgo-G\'amez \& Olofsson 2002). 
One might think that as there are less metals, the intensity of 
the lines is lower and, therefore, the ratios are also lower.  This is 
true for nitrogen but there is an anticorrelation between the 
content of oxygen and the electronic temperature (and, therefore,  the [OIII]/H$\beta$) as oxygen is a very powerful coolant (e.g. Alloin et 
al. 1979).

In order to check if the anomalous values of these two ratios in irregular 
galaxies are due to low metallicity of the parent galaxies, we plot 
the [OIII]/H$\beta$ vs. the oxygen content (Fig. 15a) and the 
[NII]/H$\alpha$ vs. the nitrogen abundances (Fig. 15b) for a sample of 
irregular galaxies: IC 10 (Hidalgo-G\'amez 2005), NGC 6822 (Hidalgo-G\'amez 
\& Peimbert 2005), DDO 53 (Hidalgo-G\'amez  \& Flores-Fajardo, in preparation), DDO50, IC 5152 (Hidalgo-G\'amez, in preparation) and 
ESO 245-G05 and Gr 8 (this work). Only five of them have reliable 
nitrogen abundances and [NII]/H$\alpha$ ratios. Three out of the five 
galaxies lie in the same line, with IC 5152  very far from this relationship. 
Gr 8 has  a very low nitrogen abundance (12+log(N/H) = 5.4; Moles et al. 1990) but the [NII]/H$\alpha$ ratio, detected at only 
one DIG location, is very high. On the contrary, DDO 53 has higher 
nitrogen abundance, $7.7$, but the line [NII]$\lambda$6583\AA~ was not 
detected at any DIG location. More data are needed to confirm the 
relationship between the nitrogen content and the [NII]/H$\alpha$ ratio.

Considering the [OIII]/H$\beta$, an inverse anticorrelation is expected 
because oxygen is an important coolant of the ISM. Then, the lower the 
oxygen content, the higher the electronic temperature is, to which the 
intensity of the [OIII]$\lambda$5007\AA~line is related.  The trend shown 
in Figure 15a is a correlation between the [OIII]/H$\beta$ and the 
oxygen content. Two of the galaxies with the higher metallicities have 
very large [OIII]/H$\beta$, while those with smaller oxygen content 
have significantly lower [OIII]/H$\beta$.  The most discordant galaxy is 
IC 5152, probably because the metal content and the DIG line ratios were 
measured in very different parts of the galaxy: the outskirts and the 
very center of the galaxy, respectively. Therefore, the 
differences in the value of the [OIII]/H$\beta$ between spiral and 
irregular galaxies are real and not due to the different metal content. 

\section{Conclusions}

In the present investigation we have studied the spectral characteristics 
of the Diffuse Ionized Gas in two irregular galaxies. In the other two 
irregular  galaxies previously studied by the author it was found that 
the [OIII]/H$\beta$ was much higher than in the DIG of spirals. One of 
the reason argued was the low metallicity   of the irregular galaxies 
(D. Cox, private communication). In order to study if this was real, we 
chose two galaxies with very low metallicity and intermediate Star 
Formation Rate.

We have studied the [OIII]/H$\beta$, the [NII]/H$\alpha$ and the 
[SII]/H$\alpha$ ratios along one slit position in ESO 245-G05. These last 
two ratios exhibit strange behaviour at the DIG locations. The main source of ionization of the DIG is photon leakage from 
the H\,{\sc  ii} regions. A similar study has been done for a total of six 
slit positions at Gr 8. Although some of the DIG data points have very 
large values of the [SII]/H$\alpha$ ratio which might indicate the presence 
of shocks, the main ionization source is, again, photon leakage with values between 40\% and 75\%.

Finally, we studied the relationship between the [OIII]/H$\beta$ and the 
oxygen content. If, as proposed, the low metallicity is the reason for 
the higher [OIII]/H$\beta$, an anticorrelation is expected between these 
two parameters. Although there are only data for seven galaxies the 
opposite trend is obtained, and therefore it can be concluded that the 
high [OIII]/H$\beta$ is not due to low metallicity.

\begin{acknowledgements}

The author is indebted to many colleages at IA-UNAM for many 
useful comments on this work. An anonymous referee is thanked for helpful and interesting comments and discussion which have improved this manuscript. She also thanks to F.J. S\'anchez-Salcedo for a carefully reading of the manuscript. This investigation is partly supported by 
CONACyT project 2002-c40366.

\end{acknowledgements}

{}




\begin{figure}
\centering
\includegraphics[width=8cm]{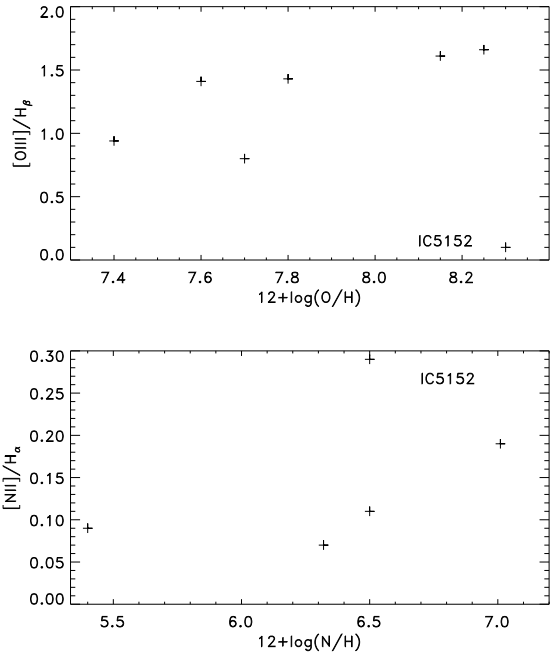}
\caption{The relation between the [OIII]/H$\beta$ of the DIG and the oxygen content of the ISM (panel a) and the ratio [NII]/H$\alpha$ and the nitrogen abundance 
(panel b) for a sample of irregular galaxies. No real correlation is found 
between them. In both panels the oddest galaxy is IC 5152. }
\label{metal}
\end{figure}

\clearpage

\begin{table}
\caption[]{ The log of the observations. 
The slit positions observed are presented in column 1. 
Column 2 shows the date of observation while the seeing is given in column 3 in arcsecond. The total integration time is presented in column 4 and the air mass in
column 5. Column 6 gives the telescope coordinates for ESO 245-G05 and the reference slit position of Gr 8 while the rest of the positions is given by the number of arcseconds the telescope drifted from the initial position. }
\vspace{0.05cm}
\begin{center}
\begin{tabular}{c c c c c c}
\hline
{Position}  & {Date}  &
{Seeing} & {Int. Time} & {Air Mas. } & {Comments} \\ 
\hline 
ESO 245-G05 No 12&  090897  & 1.2''& 90$^m$/60$^m$ & 1.08/1.04 & 01$^h$45$^m$05$^s$ -43$\degr$37$'$21$"$\\
\hline
Gr 8 slit a & 140302 &  2.0'' & 30$^m$ & 1.07 & 12$^h$58$^m$47$^s$  14$\degr$12$'$15$"$ \\
Gr 8 slit b & 140302 &  2.0'' & 30$^m$ & 1.05 & 4" to the North\\
Gr 8 slit c &  140302 & 1.5'' & 30$^m$ &1.12 & 10" to the North\\
Gr 8 slit d &  140302 & 1.5 '' & 105$^m$ &  1.3 & 13" to the North\\
Gr 8 slit e & 130302 & 1.8'' & 60$^m$ & 1.4 & 15" to the North\\
Gr 8 slit f &  130302 & 1.8'' & 45$^m$ & 1.2 & 19" to the North\\
\hline
\end{tabular}
\end{center}
\end{table}


\begin{table}
\caption[]{ The uncertainties in the lines observed for every slit position for the spectral lines studied. The name of the slit is presented in column 1, while the uncertainty, in percentage, is given in the other columns. The uncertainty in the nitrogen line is based on very few measurements for each slit.  }
\vspace{0.05cm}
\begin{center}
\begin{tabular}{c c c c c c}
\hline
{Position} & {[OIII]$\lambda$5007}  & {H$\alpha$}  &
{[NII]$\lambda$6583} & {[SII]$\lambda$6717} & {[SII]$\lambda$6731 }\\ 
\hline 
ESO 245-G05 & 27 $\%$ & 18 $\%$ & 78 $\%$ & 56 $\%$ & 55 $\%$\\
\hline
Gr 8 slit a &  7 $\%$ &  14 $\%$ & 10 $\%$ & 29 $\%$ & 23 $\%$ \\
Gr 8 slit b &  8 $\%$ &  11 $\%$ & 33 $\%$ & 18 $\%$ & 18 $\%$\\
Gr 8 slit c & 12 $\%$ &  11 $\%$ &  8 $\%$ & 50 $\%$ & 25 $\%$ \\
Gr 8 slit d &  7 $\%$ &  47 $\%$ & 52 $\%$ & 14 $\%$ & 30 $\%$\\
Gr 8 slit e &  7 $\%$ &  23 $\%$ & 19 $\%$ & 30 $\%$ & 12 $\%$ \\
Gr 8 slit f &  8 $\%$ &  11 $\%$ & -       & 11 $\%$ & 11 $\%$\\
\hline
\end{tabular}
\end{center}
\end{table}

\clearpage

\begin{table}
\caption[]{Line ratios (non-extinction corrected) of the integrated spectra for ESO 245-G05. Column 1 identifies the location along the slit. H refers to H\,{\sc ii} region and d to DIG locations. The last column corresponds to the S/N in the recombination line H$\alpha$.
}
\vspace{0.05cm}
\begin{center}
\begin{tabular}{c c c c c c c}
\hline
{Location}  & 
{ [OII]/H$\beta$} & {[OIII]/H$\beta$} & {[NII]/H$\alpha$} & {[SII]6717~\AA/H$\alpha$} & [SII]$\lambda$6717/[SII]$\lambda$6731 & S/N(H$\alpha$)\\ 
\hline 
d1 & 1.94$\pm$0.05 & 0.38$\pm$0.01   &  -  & 0.12$\pm$0.01 & 3.2 $\pm$0.3 & 5.7\\
h1 & 2.36$\pm$0.01 & 1.811$\pm$0.002 & 0.022 $\pm$0.001 & 0.057$\pm$0.002 & 1.24$\pm$0.04 & 93\\
d2 & 3.19$\pm$0.03 & -   & 0.076$\pm$0.002 & 0.174$\pm$0.006 & 1.1$\pm$0.2 & 12\\
h2 & 2.281$\pm$0.007 & 1.05$\pm$0.001 & 0.044$\pm$0.001 & 0.097$\pm$0.007 & 1.4$\pm$0.1 & 125\\
d3 & 4.1$\pm$0.1 & 0.90$\pm$0.01 & 0.075$\pm$0.004 & 0.170$\pm$0.007 & 1.61$\pm$0.07 & 10\\
h3 & 3.36$\pm$0.02 & 1.364$\pm$0.003 & 0.061$\pm$0.005 & 0.175$\pm$0.003 & 1.49$\pm$0.01 & 90\\
d4 & 6.23$\pm$0.03 & 1.16$\pm$0.001 & 0.08$\pm$0.01 & 0.196$\pm$0.004 & 1.57$\pm$0.06 & 12\\
\hline
\end{tabular}
\end{center}
\end{table}

\clearpage

\begin{table}
\caption[]{Line ratios of integrated spectra for Gr 8. Column 1 identifies 
the slit position with the letter from Table 1. H refers to H\,{\sc ii} 
regions and d to DIG locations. When more than one is at 
the same slit, a number is given.  The [OIII]/H$\beta$ is shown 
in column 2 while the [NII]/H$\alpha$ is presented in column 3, when detected. Column 4 gives the HeI/H$\beta$ for the few H\,{\sc ii} regions where it was detected and column 5 the [SII]/H$\alpha$. Finally, the extinction coefficient 
in the recombination line H$\alpha$ is shown in column 6 and the S/N in H$\alpha$ in column 7. }
\vspace{0.05cm}
\begin{center}
\begin{tabular}{c c c  c c c c}
\hline
{Location}  & 
{[OIII]$\lambda$5007\AA/H$\beta$} & {[NII]$\lambda$6584\AA/H$\alpha$} & {He I ($\lambda$5876~\AA)/H$\beta$ } & {[SII]$\lambda$6717~\AA/H$\alpha$} & C$_{\beta}$ & S/N(H$\alpha$)\\ 
\hline 
ad1 & 0.603$\pm$0.06
&  -    & -     & 0.201$\pm$0.02 & 0.11$\pm$0.006 & 5\\
ah1  & 2.693$\pm$0.1 & -     & -     & 0.057$\pm$0.004 & 2.05$\pm$0.1 & 28\\
ad2 & 0.859$\pm$0.05 & -     & -     & 0.124$\pm$0.007 & 1.01$\pm$0.05 & 7\\
bd1 & 0.520$\pm$0.03 & -     & -     & 0.272$\pm$0.02 & 0.33$\pm$0.02 & 5\\
bh1 & 2.983$\pm$0.1 & 0.016$\pm$0.001 & 0.031$\pm$0.002 & 0.061$\pm$0.004 & 2.00$\pm$0.1& 27 \\
bd2 & 1.681$\pm$0.1 &   -   &    -  & 0.074$\pm$0.01 & 1.69$\pm$0.1 & 10\\
bh2 & 1.138$\pm$0.08 &   -   &    -  & 0.074$\pm$0.005 & 2.07$\pm$0.1 & 15\\
bd3 &  -    &    -  &    -  & 0.136$\pm$0.01 & 0.65$\pm$0.1 & 5\\
cd1 & 0.995$\pm$0.09 &   -   &    -  & 0.111$\pm$0.01 & 1.22$\pm$0.06 & 2\\
ch1  & 1.402$\pm$0.07 & 0.022$\pm$0.002 & 0.022$\pm$0.002 & 0.064$\pm$0.004 & 2.11$\pm$0.1 & 32\\
cd2 & 0.342$\pm$0.04 &   -   &   -   &   -   & 1.32$\pm$0.1 & 6\\
dd1 & 1.073$\pm$0.06 &   -   &    -  & 0.155$\pm$0.01 & 1.47$\pm$0.1 & 4\\
dh1  & 1.481$\pm$0.07 &   -   & 0.012$\pm$0.0007 & 0.072$\pm$0.004 & 2.33$\pm$0.1 & 15\\
dd2 & 0.375$\pm$0.02 &   -   &    -  &   -   & 0.99$\pm$0.05 & 14\\
ed1 & 1.296$\pm$0.1 &    -  &   -   & 0.219$\pm$0.02 & 1.11$\pm$0.07 & 5\\
eh1  & 1.350$\pm$0.08 & 0.020$\pm$0.002 & 0.008$\pm$0.001 & 0.077$\pm$0.005 & 1.53$\pm$0.09 & 20\\
ed2 &   -   & 0.092$\pm$0.01 &   -   &   -   & 2.95$\pm$0.1 & -\\
fd1 & 0.719$\pm$0.1 &   -   &    -  & 0.190$\pm$0.03 & 2.04$\pm$0.2 & 2\\
fh1  & 1.122$\pm$0.06 & 0.030$\pm$0.002 &   -   & 0.078$\pm$0.005 &  1.82$\pm$0.09 & 14\\
\hline
\end{tabular}
\end{center}
\end{table}

\clearpage

\begin{table}
\caption[]{Photon leakage (in $\%$) and ionization temperatures for the  integrated spectra for Gr 8 based on the line ratios. Columns 2 and 3 correspond to the data from the excitation and columns 4 and 5 are based on the [SII]/H$\alpha$ ratio. In two cases, {\it cd1} and {\it dd1} only two values are possible.  }
\vspace{0.05cm}
\begin{center}
\begin{tabular}{c c c c c}
\hline
{Location}  & 
{Leakage [OIII] ($\%$)} & {T$_{ion}$  K} & {Leakage [SII]($\%$) } & {T$_{ion}$} \\ 
\hline 
ad1 & 58 $\%$    &  38,000  & 30-45 $\%$ & 40,000-30,000 \\
ad2 & 53 $\%$    &  38,000   & 30-40 $\%$ & 36,000-30,000  \\
bd1 & 75 $\%$    &  38,000   & 40-80 $\%$ & 38,000  \\
bd2 & 30-80 $\%$ & 47,000-38,000 & 30-35 $\%$ & 34,000-30,000  \\
bd3 &  -       &    -  & 30-40 $\%$ & 36-30,000  \\
cd1 & 50 or 80 $\%$  & 38,000 or 41,000 & 30-38  & 38,000-30,000 \\
cd2 & 55 $\%$    & 35,000    &   -   &   -   \\
dd1 & 50 or 80 $\%$ & 38,000 or 41,000 & 30-40  & 38,000-30,000 \\
dd2 &  55 $\%$   & 35,000    &    -  &   -   \\
ed1 & 40-80 $\%$ & 38,000-44,000 & 35-50 $\%$ & 40,000-30,000 \\
ed2 &   -      &    -  &   -   &   -   \\
fd1 &  60 $\%$   &   38,000 & 30-45 $\%$ & 40,000-30,000 \\
\hline
\end{tabular}
\end{center}
\end{table}

\end{document}